\documentclass[aps,floatfix,preprintnumbers,letterpaper,ifmbe,twocolumn]{revtex4}
\usepackage{graphicx,framed}
\usepackage{epsfig}
\usepackage{color}
\usepackage{amsfonts,amsmath,amsfonts,amscd}

\begin{document}

\title{Beltrami state in black-hole accretion disk: A Magnetofluid approach}

\author{Chinmoy Bhattacharjee$^{\,1,2}$, Rupam Das$^{\,3}$, David J. Stark$^{\,1,2}$,
S.M. Mahajan$^{\,1,2,4}$}

\affiliation{$^{1}$Institute for Fusion Studies, The University of
Texas at Austin, Austin,Texas 78712, USA }

\affiliation{$^{2}$Department of Physics,The University of
Texas at Austin, Austin,Texas 78712, USA}

\affiliation{$^{3}$Department of Physical and Applied Sciences, Madonna University, Livonia, Michigan 48150, USA}
\affiliation{$^{4}$Shiv Nadar University, NH91, Tehsil Dadri, Gautam Buddha Nagar, Uttar Pradesh, 201 314, India}

\begin{abstract}
Using  the magnetofluid unification  framework, we show that the accretion disk plasma (embedded in the background geometry of a blackhole) can relax to  a class of states known as the Beltrami-Bernoulli (BB) equilibria. Modeling the disk plasma as a Hall MHD system, we find that the space-time curvature can significantly alter the magnetic/velocity decay rate as we move away from the compact object; the velocity profiles in BB states, for example, deviate substantially from the predicted corresponding geodesic velocity profiles. These departures imply a rich interplay of plasma dynamics and general relativity revealed by examining the corresponding Bernoulli condition representing ``homogeneity" of total energy. The relaxed states have  their origin in the constraints provided by the two helicity invariants of Hall MHD. These helicities conspire to introduce a new oscillatory length scale into the system that is strongly influenced by relativistic and thermal effects. 
\end{abstract}

\maketitle
\section{Introduction}
Magnetized plasmas can exhibit a spontaneous tendency to evolve towards equilibrium configurations of ordered magnetic and velocity field structures. A highly investigated example of this class of equilibria is the so called ``relaxed'' state, $\nabla\times\vec{B}=\chi\vec{B}$ ($\vec{B}$ is the magnetic field configuration and $\chi$ is the Lagrange multiplier). First derived in  the context of single fluid magnetohydrodynamics (MHD) \cite{woltjer1958theorem,RevModPhys.58.741}, these states were continually extended to include two-fluid effects initially in \cite{PhysRevE.52.4287,PhysRevLett.42.1277,bhattacharjee1981energy}, and then more definitively in a later set of  papers \cite{mahajan1998double,mahajan2001formation,ohsaki2001magnetofluid,mahajan2002generation,berezhiani2015beltrami}. The essence of all these ``relaxed state'' derivations  was the construction of a constrained minimum energy principle: an appropriate energy functional was minimized while keeping the system helicities constant. The procedure results in what are known as Beltrami conditions that express the alignment of flow-vorticity along its velocity. The full equilibrium, referred to as the Beltrami-Bernoulli (BB) equilibrium, requires the simultaneous satisfaction of the  Beltrami and Bernoulli conditions, the latter signifying a homogeneous energy density. In the introduced nomenclature, the original relaxed state of Woltjer and Taylor is a single Beltrami state while the Hall MHD states are double Beltrami. The BB equilibria have been studied in relativistic and non-relativistic contexts and have yielded interesting results in several areas such as astrophysics, cosmology, high energy density physics, etc \cite{mahajan1998double,mahajan2001formation,ohsaki2001magnetofluid,mahajan2002generation,berezhiani2015beltrami}.

In this paper, we investigate an electron-ion plasma in an accretion disk near a stationary or rotating black hole. The hydrodynamics of such a plasma will be controlled by  a strong coupling between electromagnetism and gravity. We will derive the self-organized equilibrium states (for these accretion disks) by invoking the simplest non-trivial two-fluid model, Hall MHD, a two fluid formulation that neglects electron mass. Since gravity plays a dominant role near a black hole, the fluid model must be formulated in the curved spacetime geometry of general relativity (GR). As long as the structure of the curved background space-time of the embedded plasma is specified, one can incorporate  it into the special relativistic fluid model  through the   GR  extension  of the minimal coupling prescription
\begin{equation}\label{unifiedmomentum}
p^{\mu}\rightarrow A^{\mu}+\frac{m}{q}\mathcal{G}U^{\mu}=P^{\mu},
\end{equation}
invoked in the original magnetofluid unification \cite{PhysRevLett.90.035001}. \ Here, $P^{\mu}$, $A^{\mu}$, $\mathcal{G}$ and $U^{\mu}$ are the canonical 4-momenta, 4-vector potential, thermodynamic enthalpy statistical factor and plasma 4-velocity respectively. Throughout the paper, we adopt the convention $c=1$ unless explicitly stated. 

The coalescence of electromagnetic and thermodynamic attributes of the special relativistic (SR) hot fluid is achieved through the construction of an antisymmetric hybrid tensor \cite{PhysRevLett.90.035001},
\begin{equation}
M^{\mu\nu}=F^{\mu\nu}+\frac{m}{q}S^{\mu\nu},
\end{equation} obeying the equation of motion
\begin{equation}\label{MTensorSR}
qU_{\nu}{M}^{\mu\nu}= \frac{\nabla^{\mu}p-mn\nabla^{\mu}\mathcal{G}}{n}=T\nabla^{\mu}\sigma, 
\end{equation} 
where $F^{\mu\nu}$ (weight = charge q) and $S^{\mu\nu}= \nabla^{\mu} (\mathcal{G}U^{\nu})-\nabla^{\nu}( \mathcal{G}U^{\mu})$ (weight =mass m)  are the electromagnetic and composite (kinematic-statistical) fluid tensors respectively \cite{bekenstein1987helicity}; $n$ is the number density of the plasma constituents.
The right-hand side of Equation (\ref{MTensorSR}) is the thermodynamic force expressed in terms of the fluid entropy $\sigma$ and temperature $T$, found using the standard thermodynamic relation between entropy and enthalpy.

Since the advanced fluid formalism encapsulated in Eqs.~(\ref{unifiedmomentum}-\ref{MTensorSR}) looks so different from the conventional familiar MHD, a few clarifying comments that will, in particular, offer a justification for the ``language'' used in the rest of this paper, are in order:

1) In this new formalism, the Faraday tensor $F^{\mu\nu}$ (representing the electromagnetic field) and the new fluid tensor $S^{\mu\nu}$ (representing both the kinematic and the thermal attributes) appear symmetrically. In fact the tensor $M^{\mu\nu}$, just a weighted sum of $F^{\mu\nu}$ and $S^{\mu\nu}$, is an expression of the unification of the magnetic and the fluid attributes, that is, the composite system behaves as if the magnetic and fluid traits act in unison. In order to formalize this union, a new terminology was introduced in \cite{PhysRevLett.90.035001}- that of a magnetofluid. The magnetofluid, thus, is defined as a hybrid  fluid  whose ``flow''  vector is the canonical momentum $\vec{P}$ (Eq. (\ref{MTensorSR})), and its generalized vorticity (or the effective magnetic field) is $\vec{\nabla}\times \vec{P}$. Because of this deep and simplifying unification, the preceding formulation of fluid dynamics has been called the magnetofluid formalism. The usage `magnetofluid' has no connection to fluids with large permeability.

2) Even though Eq.(\ref{MTensorSR}) (with T=0, for simplicity) looks so different from the ideal MHD Ohm's law, its vector part is, in fact, nothing but a generalized Ohm's law for a magnetofluid
   \begin{equation}
   \vec{\mathcal{E}}+\vec{v}\times\vec{\mathcal{B}}=0,
     \end{equation}
where $\vec{\mathcal{E}}$ and  $\vec{\mathcal{B}}$ are simply the generalized electric and magnetic fields \cite{PhysRevLett.90.035001}. This reveals the power of the formalism, that the complicated dynamics of relativistic hot plasma has the MHD structure if expressed in suitably constructed variables. Several new results found using the formalism are given in Refs. \cite{mahajan2010twisting,pino2010relaxed,mahajan2011relativistic}.

Before returning to the generalization to curved space time, it is worthwhile to remark that the basic formalism Eqs.~(\ref{unifiedmomentum}-\ref{MTensorSR}) is fully compressible. However, later we will deal with an incompressible subclass of solutions applied to plasma in a black hole accretion disk.

Gravitational coupling changes  Eq.(\ref{MTensorSR}) to  
\begin{equation}\label{finaleom}
qU_{\nu}\mathcal{M}^{\mu\nu}=  Q T\nabla^{\mu}\sigma,
\end{equation}
where $Q=(1+\lambda f_m(R))$ with $f_m(R)$ as a function of the Ricci scalar $R$. Consequently, the new hybrid tensor $\mathcal{M}^{\mu\nu}$ defined as
\begin{equation}\label{unifiedtensordef}
	\mathcal{M}^{\mu\nu}=F^{\mu\nu}+\frac{m}{q}D^{\mu\nu}
\end{equation}
with 
\begin{equation}\label{unifiedtensordef}
	D^{\mu\nu}=(1+\lambda f_{m}(R) -\lambda RF_m)S^{\mu\nu}+\frac{m}{q}\lambda F_mK^{\mu\nu}
\end{equation}
includes an additional gravity coupled flow field tensor, $K^{\mu\nu}=\nabla^{\mu}(R\mathcal{G}U^{\nu})-\nabla^{\nu}(R\mathcal{G}U^{\mu})$\cite{bhattacharjee2015magnetofluid}. Here, $F_{m}(R)=f_{m}'(R)$ and $\lambda$ is the phenomenological parameter representing the coupling between matter and background geometry \cite{bhattacharjee2015magnetofluid,balakin2005non,balakin2014nonminimal,Harko:2008qz}.

With this new formalism, the Beltrami-Bernoulli conditions of interest can be found by analyzing the new 3D vorticity evolution equations obtained by  performing a $3+1$ decomposition (of spacetime) of Eq. (\ref{finaleom}). We find that, working within the framework  of the Hall MHD model, the  equilibrium of the gravitating accretion disk plasma, just like in its flat space-time counterpart, is described by a double Beltrami state. Our derived fluid velocity structure (and the corresponding magnetic field profile) in both geometries exhibits a faster radial decay compared to that in geodesic motion; the change reflects the electromagnetic interactions present in the fluid. Furthermore, the two-fluid configuration of our system introduces radial oscillatory structure to the velocity and magnetic field that is dependent upon the new generalized helicities.

The paper is arranged as follows: We first present a brief overview of the vorticity evolution equation, obtained by applying the Arnowitt-Deser-Misner (ADM) formalism of electrodynamics \cite{MTW,thorne1982electrodynamics,Wald,thorne1986black,bhattacharjee2015magnetofluid} (Appendix \ref{magnetofluiddynamics}) to Eq. (\ref{finaleom}), the  magnetofluid equation of motion in curved space-time. Then, we derive the equations for the so called relaxed state equilibrium. After displaying an analytical solution of the equilibrium configuration in the non-relativistic case, we adopt numerical methods to solve the more complicated  General Relativistic equilibrium. Finally, we compare the results obtained in the classical and General Relativistic cases by deriving the corresponding Bernoulli condition to explore the new ``structure'' induced by GR.

\section{Vortical dynamics}\label{vorticaldynamics}
In the 3+1 decomposition (see Appendix \ref{magnetofluiddynamics}), the derivation of the 3D vorticity evolution equation begins with taking  the spacelike projection (the application of the spacelike projection operator $\gamma^{\beta}\ _{\mu}$) of the unified field equation of motion (\ref{finaleom}). The resulting  momentum evolution is
\begin{equation}\label{generalmomeq}
\alpha q \Gamma \vec{\xi}+q\Gamma(\vec{v}\times\vec{\Omega})=-(1+\lambda f_{m}(R))T\vec{\nabla}\sigma.
\end{equation}
Similarly, the timelike projection (the application of $n_{\mu}$) gives  the equation of energy conservation
\begin{equation}\label{generalenergyeq}
\alpha q\Gamma \vec{v}\cdot\vec{\xi}=T (1+\lambda f_{m}(R))(\mathcal{L}_t\sigma-\vec{\beta}\cdot\vec{\nabla}\sigma),
\end{equation}
where $\vec{\xi}$ and $\vec{\Omega}$ are, respectively, the generalized electric and magnetic fields [ Eqs. (\ref{generalefield}) and (\ref{generalbfield})]. Also, $\alpha$, $\Gamma$, and $\vec{v}$ (spatial velocity of plasma) are defined in Eqs. (\ref{canonicalmetric}), (\ref{lorentzfactor}), and (\ref{fourvelocity}). Here,  $\alpha$ represents the lapse function needed to capture the change of proper time between two adjacent spacelike hypersurfaces, and $\Gamma$ represents the Lorentz factor in curved spacetime.

Next, the desired vorticity evolution equation (which is really the generalized Faraday's law) with gravity driven sources  can be obtained from the momentum evolution equation Eq. (\ref{generalmomeq}) by using the antisymmetric property of $\mathcal{M}^{\mu\nu}$, $\nabla_{\mu}\mathcal{M}^{*\mu\nu}=0$. Combining  the $\gamma^{\beta}\ _{\mu}$  projection of the preceding identity, 
\begin{equation}\label{faradaygeneral}
\mathcal{L}_t\vec{\Omega}=\mathcal{L}_{\vec{\beta}}\vec{\Omega}-\vec{\nabla}\times(\alpha\vec{\xi})-\alpha \Theta \vec{\Omega},
\end{equation}
with the momentum evolution equation (\ref{generalmomeq}), gives us the generalized vorticity evolution equation \cite{asenjo2013generating,bhattacharjee2015magnetofluid,bhattacharjee2015novel}
\begin{equation}\label{vorticity}
\mathcal{L}_t\vec{\Omega}-\vec{\nabla} \times (\vec{v}\times \vec{\Omega})-\mathcal{L}_{\vec{\beta}}\vec{\Omega}+\alpha \Theta \vec{\Omega}=\vec{\nabla}\times\left(\frac{T}{q\Gamma}(1+\lambda f_{m}(R))\vec{\nabla}\sigma\right),
\end{equation}
where $\mathcal{L}$  denotes Lie derivatives with $\mathcal{L}_t = \partial_t$ along $t^{\mu}$,  $\mathcal{L}_{\vec{\beta}}{\vec{\Omega}}= [\vec{\beta},\vec{\Omega}]$, and the expansion factor $\Theta$ is defined in Appendix \ref{magnetofluiddynamics}. The shift function $\beta$ denotes the motion on spacelike hypersurfaces during the evolution of 3D surfaces, and $\Theta$ measures the average expansion of the infinitesimally nearby surrounding geodesics (or congruences). It should be noted that, in this formalism, the plasma is coupled to gravity even in the limit $\lambda=0$.

One cannot fail  to observe that all terms on the left-hand side of the generalized vorticity evolution equation (\ref{vorticity}) operate on the vorticity 3-vector $\vec{\Omega}$, while the right-hand side  provides, just as in the conventional picture, possible sources for generating  $\vec{\Omega}$.  
The left-hand side, however, has considerably more structure than the conventional 3D vortex dynamics; the first two terms reflect the standard Helmholtz vortical dynamics 
whereas $\alpha \Theta \vec{\Omega}$ and $\mathcal{L}_{\vec{\beta}}\vec{\Omega}$ are nontrivial gravity modifications. Thus, the gravity coupling does, fundamentally, modify the projected 3D vortex dynamics, in spite of the fact that the 4D vortex equations had exactly the same form and that this modification was manifest only through the non-inertial splitting of spacetime.

\subsection{Equilibrium states} 

In this paper, we will investigate the equilibrium states in two background geometries, i.e., `Schwarzschild' and `Kerr' with minimal coupling only.  The standard metric describing the stationary and axially symmetric  (static and spherically symmetric) spacetime for Kerr (Schwarzschild) black holes can be written as \cite{Harkolobo}
\begin{equation}\label{ASmetric}
	ds^2=g_{tt}dt^2+2g_{t\phi}dtd\phi+g_{rr}dr^2+g_{\theta\theta}d\theta^2+g_{\phi\phi}d\phi^2.
\end{equation}

In the geometries described by the above metric, it can be shown that the terms involving $\Theta$ and $\beta$ in Eq. (\ref{vorticity}) vanish and, therefore, make no contribution to the vortical dynamics of the plasma. While the vanishing of $\Theta$ follows directly from the metric, the vanishing of the $\beta$ term follows from the assumption that the generalized vorticity $\Omega$ varies along the radial direction only. 

We will concentrate, in this paper, on the simplest vortical dynamics without any sources (originating in inhomogeneous thermodynamics or in nontrivial-gravity modifications), i.e, the left hand side of Eq. (\ref{vorticity}) is put equal to zero,
\begin{equation}\label{simple_vor}
\mathcal{L}_t\vec{\Omega}-\vec{\nabla} \times (\vec{v}\times \vec{\Omega})=0.
\end{equation} 
The source free vortical dynamics allows the conservation of helicity, $\mathcal{H} = <\vec{\Omega}\cdot {\nabla\times}^{-1}\vec{\Omega}>$, constructed out of the generalized vortical field and its corresponding canonical momenta. The next task is to  work out possible equilibrium solutions of the aforementioned vortex dynamics relevant to an accretion disk plasma consisting of electrons and ions; each species has its own vortical dynamics- vorticity, helicity, etc.   

We will not consider here a very special, well-known superconducting solution $\vec{\Omega}=0$ of Eq. (\ref{simple_vor}) since it yields a rather trivial solution for the accretion disk\cite{mahajan2008classical}. Instead, we will investigate the equilibrium relaxed solution $\vec{\Omega}=\Lambda\vec{v}$ that makes the generalized force vanish by aligning the vorticity and the flow (Beltrami condition). The separation constant $\Lambda$ is equivalent to a Lagrange multiplier if we were to derive the Beltrami condition by the standard constrained energy minimization procedure.  The  equilibrium is fully determined by the two Beltrami conditions (one for each species), and Ampere's law:

\begin{eqnarray}
	\vec{B}+\frac{m_ec}{q}(\nabla\times(\mathcal{G}\Gamma_e\vec{v_e}))=\Lambda_e\vec{v_e}\label{elec1},\\
	\vec{B}+\frac{m_ic}{q}(\nabla\times(\mathcal{G}\Gamma_i\vec{v_i}))=\Lambda_i\vec{v_i}\label{ion1},\\
	\nabla\times(\sqrt{-g_{tt}} \ \vec{B})=\frac{4\pi\sqrt{-g_{tt}}\ q n(r)}{c}(\Gamma_i\vec{v_i}-\Gamma_e\vec{v_e})\label{amph1},
\end{eqnarray} 
where $n(r)$ is the density profile of the fluid in the accretion disk. The separation constants $\Lambda_{e,i}$ are Lagrange multipliers that are to be, eventually, related to the two topological constants - the conserved individual helicities of the two species. These two helicities, as constants of motion, are two labels of an equilibrium system. Note that the 3-gradient in the above equations is also modified through spacetime metric elements; for an arbitrary scalar function $P$, it can be written as 
\begin{equation}\label{gradscalar1}
	\vec{\nabla} P=\frac{1}{\sqrt{g_{rr}}}\partial_rP \ \hat{e}_r+\frac{1}{\sqrt{g_{\theta\theta}}}\partial_{\theta}P\ \hat{e}_{\theta}+\frac{1}{\sqrt{g_{\phi\phi}}}\partial_{\phi}P \ \hat{e}_{\phi}.
\end{equation}

\section{Non-Relativistic Solution}
An analytic investigation of Eqs. (\ref{elec1}-\ref{amph1}) is quite impossible because they are highly nonlinear in the velocities. However, to extract some basic features of the equilibrium states, let us analyze the system in its non-relativistic limit:  $\Gamma\rightarrow 1$, thermodynamic statistical factor $\mathcal{G}\rightarrow 1$, $\sqrt{- g_{tt}}\rightarrow 1$ and $m_e << m_i$.  The resulting normalized equations are:
\begin{eqnarray}
\vec{B}=\mu_e\vec{v}_e \label{elec2}, \\
\vec{B}+\nabla\times\vec{v}_i=\mu_i\vec{v}_i \label{ion2} ,\\
\nabla\times\vec{B}= \eta\hat{n}(\vec{v}_i-\vec{v}_e) \label{amph2},
\end{eqnarray} 
where all the gradients are normalized to a macroscopic length scale $r_g=GM/c^2$ (half of the Schwarzschild radius used below), velocities are normalized to the speed of light $c$, $\eta=r_g^2/\lambda_s^2$ is a dimensionless factor, $\hat{n}(r)$ is the density factor, $q$ is the ion charge, $\lambda_s^2=c^2/\omega_p^2$ is the ion skin depth, $\omega_p^2=(4\pi q^2n_0)/m_i$ is the ion-plasma frequency and $\mu_{e,i}=(qr_g/m_ic)\Lambda_{e,i}$. Notice that all fields are divergence-free (solenoidal), so our analysis is localized to an incompressible subclass of solutions.

Manipulating Eqs. (\ref{elec2})-(\ref{amph2}), 
we derive a single second order differential equation 
\begin{eqnarray}\label{eq1}
-\mu_e\nabla^2\vec{v}_e+(\eta\hat{n}-\mu_i\mu_e)\vec{\nabla}\times\vec{v_e}=\nonumber \\
\eta\hat{n}(\mu_i-\mu_e)\vec{v}_e-\mu_{e}\hat{n}(\vec{\nabla}\hat{n}^{-1})\times(\vec{\nabla}\times\vec{v_e})
\end{eqnarray}
for the velocity $\vec{v}_e$.  For spherically symmetric solutions for the solenoidal fields, the radial component $v_r=0$. The equations for $v_{\theta}(r)$ and $v_{\phi}(r)$ are, then, written as 
\begin{eqnarray}
\frac{\mu_e}{r^2}\frac{\partial}{\partial r}\left(r^2\frac{\partial}{\partial r}v_{\theta}\right)-\mu_{e}\frac{v_{\theta}}{r^2}+\eta\hat{n}(\mu_i-\mu_e)v_{\theta}=\nonumber \\
(\mu_i\mu_e-\eta\hat{n})\frac{1}{r}\frac{\partial(rv_{\phi})}{\partial r}+\frac{\mu_e}{r}\frac{\partial (\log\hat{n})}{\partial r}\frac{\partial (rv_{\theta})}{\partial r}\label{theta1},\\
\frac{\mu_e}{r^2}\frac{\partial}{\partial r}\left(r^2\frac{\partial}{\partial r}v_{\phi}\right)-\mu_{e}\frac{v_{\phi}}{r^2}+\eta\hat{n}(\mu_i-\mu_e)v_{\phi}=\nonumber \\
(\eta\hat{n}-\mu_i\mu_e)\frac{1}{r}\frac{\partial (rv_{\theta})}{\partial r}+\frac{\mu_e}{r}\frac{\partial (\log\hat{n})}{\partial r}\frac{\partial(rv_{\phi})}{\partial r}\label{phi1}.
\end{eqnarray}
Here, we dropped the index  $e$ from $\vec{v}_e$ for simplicity.

In terms of the variables, $Q_{\theta}=rv_{\theta}$ and $Q_{\phi}=rv_{\phi}$, we find autonomous (in r) equations   
\begin{eqnarray}
\mu_e\frac{\partial^2 Q_{\theta}}{\partial^2 r}-\mu_e\frac{Q_{\theta}}{r^2}+\eta\hat{n}(\mu_i-\mu_e)Q_{\theta}=\nonumber \\
(\mu_i\mu_e-\eta\hat{n})\frac{\partial Q_{\phi}}{\partial r}+\mu_e\frac{\partial (\log\hat{n})}{\partial r}\frac{\partial Q_{\theta}}{\partial r},\label{theta2}\\
\mu_e\frac{\partial^2Q_{\phi}}{\partial^2 r}-\mu_e\frac{Q_{\phi}}{r^2}+\eta\hat{n}(\mu_i-\mu_e)Q_{\phi}=\nonumber \\
(\eta\hat{n}-\mu_i\mu_e)\frac{\partial Q_{\theta}}{\partial r}+\mu_e\frac{\partial (\log\hat{n})}{\partial r}\frac{\partial Q_{\phi}}{\partial r}\label{phi2}
\end{eqnarray} 
that are readily solved in terms of exponentials: $Q_{\theta}=\hat{Q}_{\theta}e^{sr}$ and $Q_{\phi}=\hat{Q}_{\phi}e^{sr}$ where the exponent $s$ is determined by the quadratic
\begin{equation}\label{disp}
\mu_e s^2-\frac{\mu_e}{r^2}+\eta\hat{n}(\mu_i-\mu_e)-\mu_e s\left(\frac{\partial (\log\hat{n})}{\partial r}\right)=\pm i s(\eta\hat{n}-\mu_i\mu_e).
\end{equation}

One must recognize that the parameter $\eta$, the ratio of the Schwarzschild radius to the ion skin depth, is $\gg 1$. Thus, the dominant balance in Eq.(\ref{disp}) yields
\begin{equation}\label{disp}
 s \simeq\mp i(\mu_i-\mu_e).
\end{equation}
In this limit $v_{e}=v_{i}$, and  the azimuthal and poloidal velocities are
 \begin{eqnarray}\label{solution}
v_{\theta}=\frac{C}{r}\cos[(\mu_i-\mu_e)r+\delta],\\
v_{\phi}=-\frac{C}{r}\sin[(\mu_i-\mu_e)r+\delta],
\end{eqnarray}
where $C$ and $\delta$ are constants determined by the initial conditions. It should be noted here that, to this order, the solutions are independent of density profiles.
\begin{figure}
	\centering
	\includegraphics[width=0.95\linewidth]{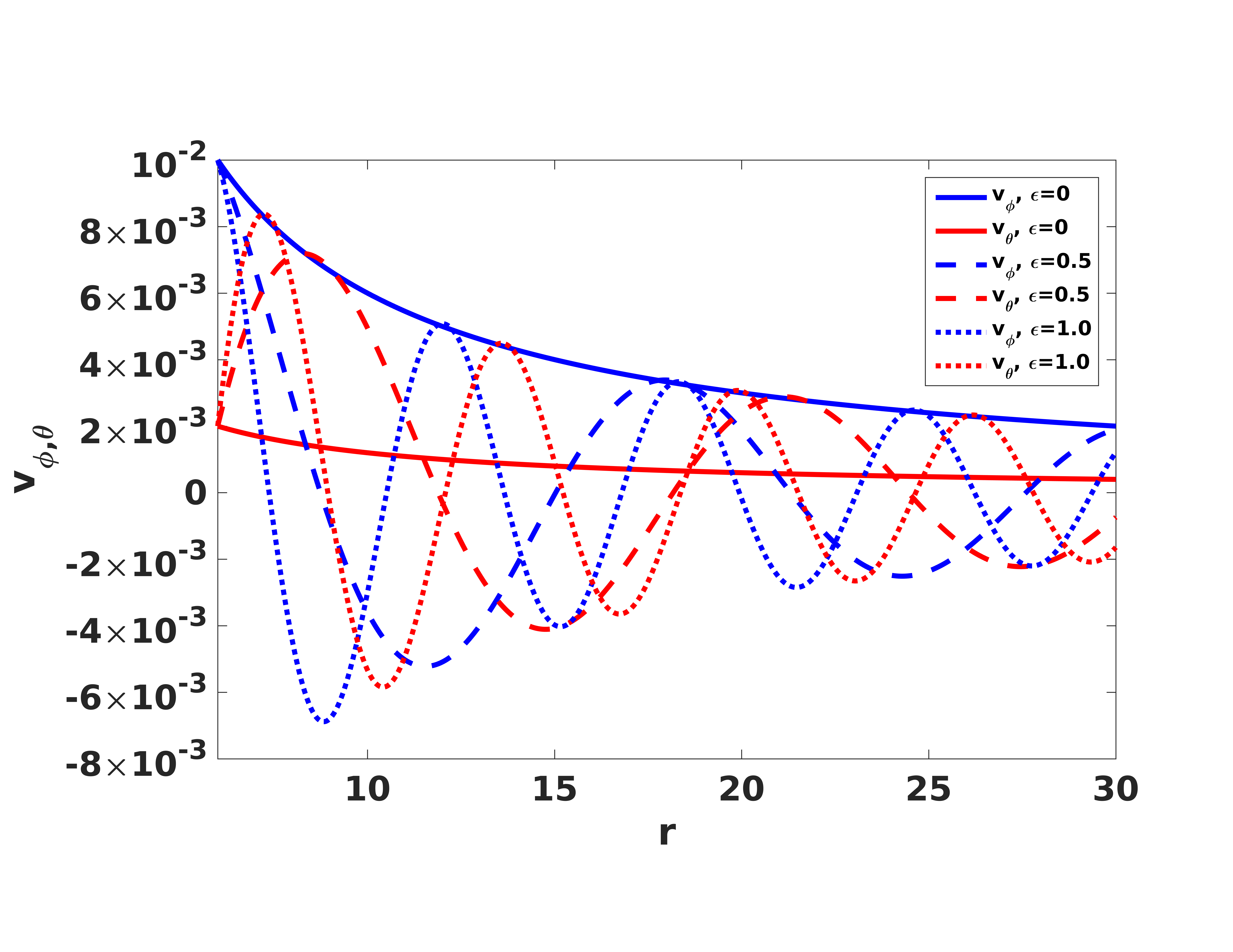}
	\caption{(Color Online)Velocity profiles for the non-relativistic solution with different values of $\epsilon=\mu_i-\mu_e$. The blue (upper) and red (lower) solid lines represent the $v_{\phi}$ and $v_{\theta}$, respectively, for $\epsilon=0$. The dashed and dotted curves represent the velocity components for the $\epsilon=0.5$ and $\epsilon=1$ cases, respectively. The blue (initially at $10^{-2}$) curves represent $v_{\phi}$, whereas the red curves (initially at $2\times10^{-3}$) curves denote $v_{\theta}$. As the $\eta$ values change, the frequency of the oscillating solution varies.}
	\label{fig:classical}
\end{figure}
In all figures displayed in this paper, we will plot only the above equilibrium velocity profiles. Since the magnetic field is simply proportional to velocity, the plots could just as well be seen as representing normalized magnetic fields.

In Fig. (\ref{fig:classical}), we plot $v_\theta$ and $v_\phi$  for different values of $\epsilon=\mu_i-\mu_e$.
When the two Lagrange multipliers are the same, one observes only  the  $\sim r^{-1}$ decay. Oscillations ensue for nonzero $\epsilon$, increasing in frequency as $\epsilon$  becomes larger. The  relatively slow $\sim r^{-1}$ decay is overwhelmed by radial oscillations on a new scale dictated by $\epsilon$. The origin of this scale  lies in the values of the system invariants that  manifest through the Lagrange multipliers; it is different, for instance, from the intrinsic scale defined by the ion skin depth. 

The preceding non-relativistic solutions, in the ($\eta\rightarrow\infty$) limit, provide us a fiducial backdrop for the study of more advanced/numerical equilibrium solutions of accretion disk plasmas embedded in curved space-time. Before proceeding in that direction,  it is easy to calculate the first order  correction to the frequency:
\begin{eqnarray}
s_1(r)=\pm\frac{i\mu_e}{\eta\hat{n}}\left[(\mu_i-\mu_e)^2-\mu_i(\mu_i-\mu_e)+\frac{1}{r^2}\right]\nonumber \\
+\frac{\mu_e(\mu_i-\mu_e)}{\eta\hat{n}}\partial_r\log{\hat{n}}.
\end{eqnarray}  
Note that for $\mu_i=\mu_e$, there are still oscillations to this order and the solution now depends on the density profiles.

\section{General Relativistic Solution} \label{GR}
In General Relativity, the nonlinearity of Eqs. (\ref{elec1}-\ref{amph1}) describing the  equilibrium state of a hot fluid-like plasma (minimally coupled to gravity) becomes exacerbated  with gravity entering into the equations through both spacetime metric elements and the Lorentz factor. While seeking the equilibrium state of a plasma in the accretion disk around a black hole, one must remember that the accreting plasma gains kinetic energy but loses angular momentum as it approaches the black hole \cite{vietri2008foundations}. An analysis of the dynamics of plasma in these disks is crucial for obtaining its equilibrium state.

In order to obtain the equilibrium states of plasma in accretion disks, it is necessary to ``choose'' the correct velocity dependence of the associated Lorentz factor $\Gamma$ buried in the equations (\ref{elec1}-\ref{amph1}). The correct choice can be made only after the following observations. First, we assume that the plasma itself does not contribute to the background spacetime structure specified by the spacetime metric (\ref{ASmetric}) and, therefore, the plasma behaves like test matter moving in the curved spacetime of the black hole. Thus, the plasma particles would follow the geodesic motion (dictated by the spacetime metric) if the electromagnetic interactions were absent. In other words, these plasma particles with mass $m$, charge $q$ and four velocity $U^{\mu}$ in geodesic motion will satisfy $a^{\nu}=U^{\mu}\nabla_{\mu}U^{\nu}=0$ ($a^{\nu}$ is the absolute acceleration of the plasma particles defined in Appendix \ref{magnetofluiddynamics}). However, when electromagnetism is turned on, the particle trajectories will obey $a^{\nu}=U^{\mu}\nabla_{\mu}U^{\nu}=(q/m)F^{\nu}_{\ \ \mu}U^{\mu}$.  One must, therefore, be cognizant of the  non-geodesic motion induced by an  external electromagnetic field. 

Second, the magnetofluid formalism provides an elegant approach to determining the velocities of plasma elements interacting electromagnetically. For minimal coupling to gravity, it easily follows from Eq. (\ref{finaleom}) that the force-free motion of plasma particles with electromagnetic interactions can be described by $qU_{\nu}\mathcal{M}^{\mu\nu}=0$, which is equivalent to $U^{\mu}\nabla_{\mu}U^{\nu}=(q/m)F^{\nu}_{\ \ \mu}U^{\mu}$, for $\mathcal{G}\rightarrow 1$  (i.e., by turning off thermodynamics), when the latter is cast in unified magnetofluid variables. The major difference is that the former can also describe the dynamics of a continuous matter distribution (with the inclusion of thermodynamics) like plasma in an equilibrium state, whereas the latter describes the dynamics of only discrete charged particles in an external electromagnetic field. 

Third, with a 3+1 decomposition of spacetime, the force-free (equilibrium) motion of plasma particles in an external electromagnetic field can also be described by Eqs. (\ref{elec1}-\ref{amph1}) for $\mathcal{G}\rightarrow 1$ and $\Gamma = -(1/\alpha) n_{\mu} U^{\mu}={1}/{\sqrt{-g_{tt}-v^2}}$ as defined in Appendix \ref{magnetofluiddynamics}. Note that the second equality follows from $U^{\mu}U_{\mu} = -1$. Physically, the Lorentz factor $\Gamma$ relates an Eulerian observer (flowing with the timelike normal vector $n^\mu$ perpendicular to the spacelike hypersurface) to a Lagrangian observer moving with plasma four-velocity $U^\mu$ at each point in spacetime. Since the plasma four-velocity $U^\mu$ at each point in the accretion disk is the consequence of all the forces (not just gravity) acting on the plasma, it can be used in $\Gamma$ in equations (\ref{elec1}-\ref{amph1}) to solve for the equilibrium states of plasma. Thus, for geodesic motion, $\Gamma$ is completely determined by the corresponding metric elements, whereas it is determined from both metric elements and velocities in non-geodesic motion of equilibrium state plasma, the latter of which need to be determined by Eqs. (\ref{elec1}-\ref{amph1}). 

Finally, for geodesic motion, the plasmas rotate along stable circular orbits around black holes with $v_{\theta}\approx 0$. In the thin accretion disk model, it is also assumed that $v_{\phi}>>v_{r}$. However, since we investigate a system of plasma influenced by both gravity and electromagnetism, we relax the condition of a negligible poloidal velocity $v_{\theta}$ of a thin accretion disk and search for the equilibrium magnetic/velocity field profile in the equatorial plane of the accretion disk, i.e., at $\theta=\pi/2$. 

Using the Lorentz factor
\begin{align}\label{mod_lorentz}
\Gamma=\frac{1}{\sqrt{-g_{tt}-v_{\theta}^2-v_{\phi}^2}},
\end{align}
we will now numerically solve Eqs.(\ref{elec1}-\ref{amph1}) to obtain the equilibrium states of plasma in accretion disks of both Schwarzschild and Kerr geometry.

\subsection{Schwarzschild geometry} \label{SC}

For a spherically symmetric and static space-time (Schwarzschild space-time), the relevant space-time metric elements are 

\begin{alignat}{2}\label{smetric}
g_{tt}=-\alpha^2=-(1-\frac{2r_g}{r})\ ;\  &\qquad\text{} g_{rr}=(1-\frac{2r_g}{r})^{-1},\notag \\
g_{\theta\theta}=r^2\ ;\  &\qquad\text{}  g_{\phi\phi}=r^2\sin^2\theta.
\end{alignat}

For this geometry, normalized Eqs. (\ref{elec2})-(\ref{amph2}) read
 
 \begin{eqnarray}
\vec{B}=\mu_e\vec{v}_e \label{elec3}, \\
\vec{B}+(\nabla\times(\mathcal{G}\Gamma_i\vec{v}_i))=\mu_i\vec{v}_i, \label{ion3} \\
\nabla\times (\alpha \vec{B})= \eta\hat{n}(r)\alpha(\Gamma_i\vec{v}_i-\Gamma_e\vec{v}_e) \label{amph3},
\end{eqnarray}
where the density factor $\hat{n}$ can be a function of radial distance in the accretion disk, decaying as we move away from the event horizon. We have numerically solved this system again in the overwhelmingly justified limit $\eta \rightarrow \infty$. 

\begin{figure}
\centering
\includegraphics[width=0.95\linewidth]{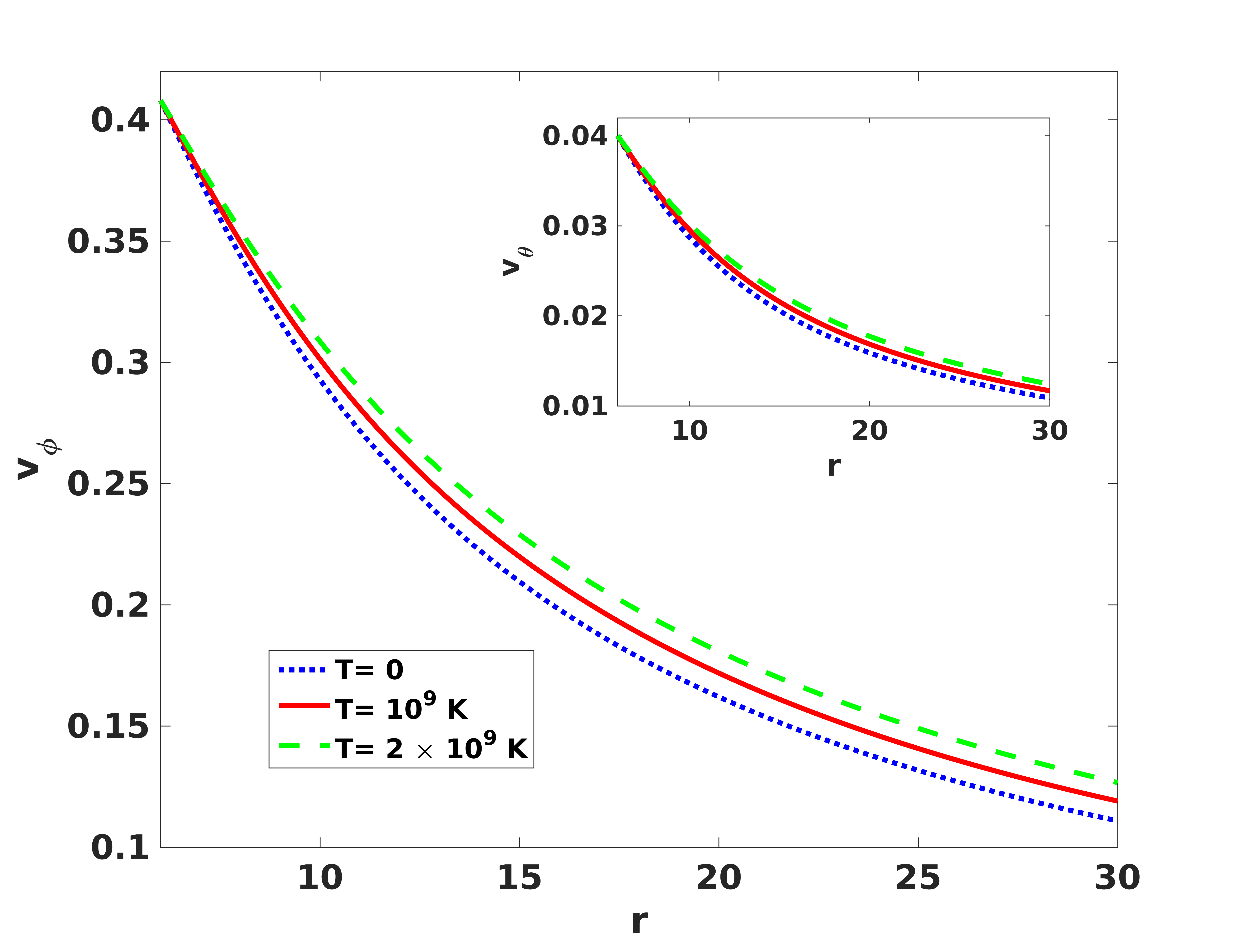}
\caption{(Color Online)Decaying solution of $v_{\phi}$ and $v_{\theta}$ in an accretion disk near a Schwarzschild black hole for initial values of 0.408 and 0.04 respectively.  Starting from $r=6$, three lines for both velocity profiles show decay for three different initial values of temperature from $T=0K$ (cold) to $T=2\times 10^9 K$(hot). For both velocity profiles, the decaying solution with the highest temperature decays slowest.}
\label{fig:Scharz_decay}
\end{figure}

Fig. (\ref{fig:Scharz_decay}) shows the velocity profiles for $v_{\phi}$ and $v_{\theta}$. The main figure gives the profiles for the azimuthal velocity $v_{\phi}$,  whereas the inset shows the poloidal velocity $v_{\theta}$ for three different values of the thermodynamic function $\mathcal{G}$. This factor is approximated as $\mathcal{G}\approx 1+(5/2) \ k_bT/mc^2$ with $k_b$ being the Boltzmann constant \cite{PhysRevLett.90.035001}. The temperature profile is chosen to be that of a blackbody, $T(r)\sim r^{-3/4}$. It should be noted that in standard accretion disk theory with proper GR corrections, the accretion disk is divided into inner, middle and outer regions with slightly different temperature profiles for each region \cite{shakura1989black,novikov1973astrophysics}. Since the goal of the paper is to investigate the changes due to GR, we adopt only one profile and note that the qualitative features of our results are still valid for different regions in the accretion disk.

Starting with initial values of $v_{\phi}(r=6)=0.408$ and $v_{\theta}(r=6)=0.04$ for a system with $\epsilon=0$, the plot shows the decay rate for both velocity components slowing down with increasing temperature.  It should be noted here that our system has a certain ambiguity for the exact initial values of velocity profiles, as we can set the values either far away or at the inner most stable circular orbit (isco) in the black hole accretion disk. Unlike geodesic motion where the azimuthal velocity for the isco, $v_{\phi}(r=6)=0.408$,  can be derived solely from the metric elements and their radial derivatives as shown in Eq. (\ref{vgeodesic}), determining the same for non-geodesic motion requires solving the involved orbital plasma dynamics in accretion disks with electromagnetic interactions as well as a complicated set of disk structure equations in GR. On the other hand, as $r\rightarrow \infty$, it is obvious that geodesic and non-geodesic velocities approach zero which, though a viable
analytical option for an initial value, turns out be infeasible for numerical computation. Therefore, initial values for geodesic velocities at isco in the accretion disk can serve as a convenient benchmark for numerical computation; these in fact have been used in this paper for both black hole geometries.

\begin{figure}
\centering
\includegraphics[width=0.95\linewidth]{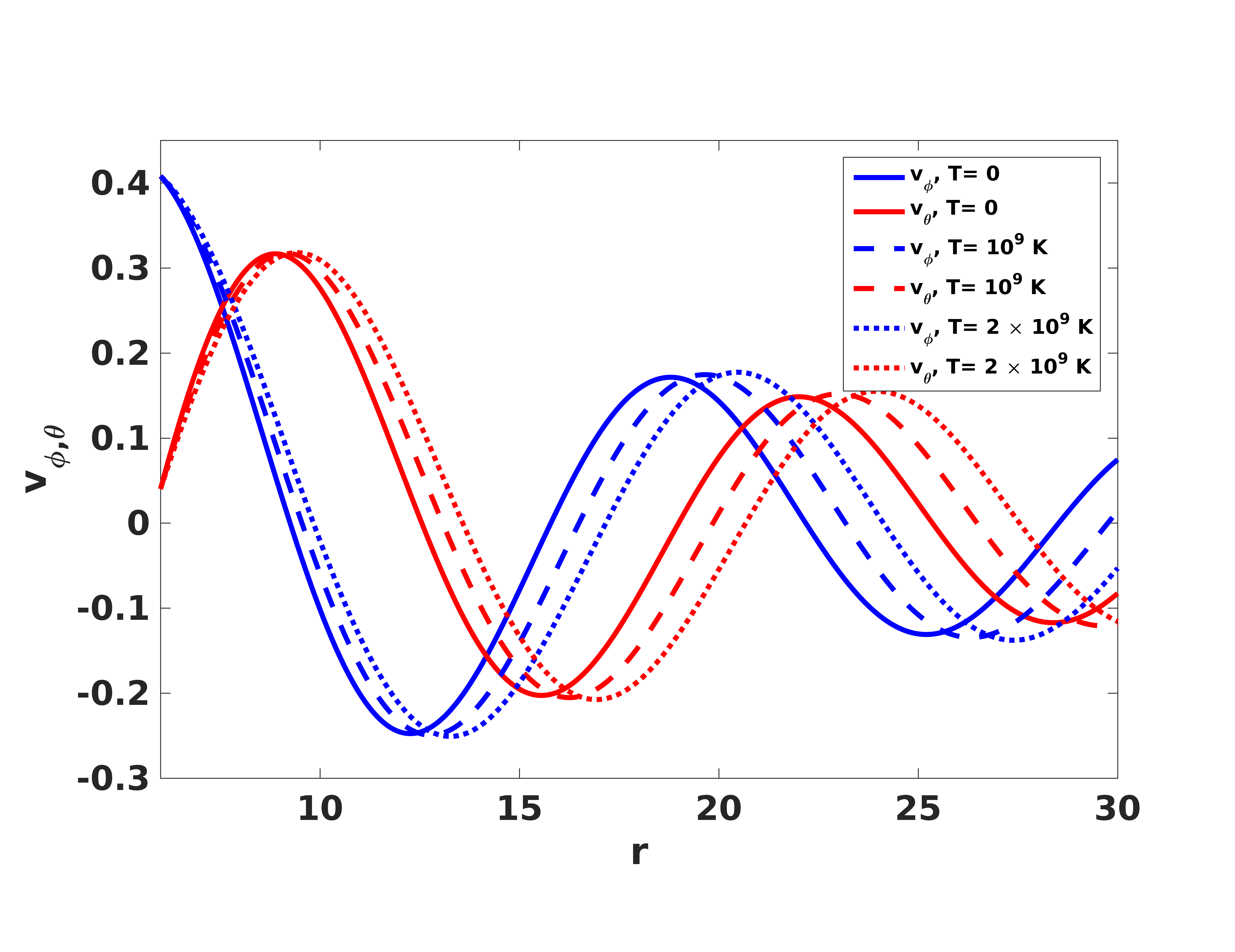}
\caption{(Color Online)Oscillatory solution of $v_{\phi}$ and $v_{\theta}$ in an accretion disk near a Schwarzschild black hole for initial values of 0.408 and 0.04 respectively. Starting from $r=6$, three oscillatory lines for both velocity profiles show decay for different initial temperatures $T=0K$ (solid), $T=10^9 K$ (dashed), and $T=2\times 10^9 K$(dotted). Blue lines (initially upper) and red lines (initially lower) represent $v_{\phi}$ and $v_{\theta}$ respectively. For both velocity profiles, the frequency of oscillation changes as the temperature changes.}
\label{fig:compareS_osc}
\end{figure}

Fig.(\ref{fig:compareS_osc}), on the other hand, shows an oscillatory decaying profile of both velocity components in the accretion disk near a Schwarzschild black hole. This oscillatory behavior can be understood from the classical solution where sinusoidal behavior emerges as soon as there is a difference between the two Lagrange multipliers, in this case $\epsilon=0.5$. The difference in Lagrange multipliers introduces a new length scale on which the new oscillatory behavior depends. In contrast to the decaying solution, these profiles show change in frequency as well as decay rate for different temperatures. Similar to the classical case, the poloidal velocity oscillates approximately $\pi/2$ out of phase and with the same decaying amplitude as the azimuthal velocity. 

As a simple exercise to gain insight into the role gravity plays in shaping the equilibrium states, we explore the regime where $v^2<<r^{-1} << 1$ and Taylor expand $\alpha$ and $\Gamma$ to lowest order: $\alpha \approx 1 - 1/r$ and $\Gamma \approx 1 + 1/r $. The approximate solutions, now,  still take the form  $Q_{\theta}=\hat{Q}_{\theta}e^{s(r)r}$ and $Q_{\phi}=\hat{Q}_{\phi}e^{s(r)r}$, where  $s(r)$ obeys

\begin{equation}\label{disp_app}
\left(1-\frac{1}{r}\right)\left[\left(s+r\frac{ds}{dr}\right)\left(1+\frac{1}{r}\right)-\frac{1}{r^2}\right]=\pm i (\mu_i-\mu_e),
\end{equation}
and is solved as
\begin{equation}\label{app_sol}
s(r)=\pm i (\mu_i-\mu_e)\left(1-\frac{\tanh r}{r}\right)+\frac{1}{r}\log\left(\frac{r}{1+r}\right),
\end{equation}
implying that the frequency and the decay rate have changed compared to the classical solution; the changes are most pronounced at small $r$. In particular, now $v\sim 1/(1+r)$ so that the decay rate is slower than the $1/r$ decay seen in the classical case; as expected, it approaches $1/r$ for large $r$. Furthermore, the frequency of oscillation also reduces at small distances from the black hole due to the additional $\tanh(r)$ term. In other words, the curvature of space-time tends to extend the length scales associated with the variations in the velocity profiles.

\subsection{Kerr geometry}
Now, for an axisymmetric (but not spherically symmetric) stationary system, like the Kerr black hole, the shift function $\beta$ is non-zero. Consequently, the pertinent  equation (\ref{vorticity}), in general,  does not show any similarity to standard 3D vortex dynamics. The shift function for rotating black holes can be taken to be $\vec{\beta}=-\omega \ \hat{e}_{\phi}$, where $\omega = -g_{t\phi}/g_{\phi\phi}$ represents the angular velocity of the black hole measured by a zero angular momentum observer. The term involving the shift function, however,  will give zero contribution since we assumed that $\Omega$  has only radial dependence. It can be further shown that the term involving the expansion factor $\Theta$ vanishes for the Kerr metric as well. Thus, the solution $\vec{\Omega}=\Lambda\vec{v}$ is valid in this case also. 

The relevant space-time metric elements in Boyer-Lindquist coordinates are \cite{2013A&A551A4P}
\begin{alignat}{2}\label{kerrmetric1}
g_{tt}=-\alpha^2=-\frac{(\Delta_r-a^2)}{r^2}\ ;\  &\qquad\text{} g_{rr}=\frac{r^2}{\Delta_r},\notag \\
g_{t\phi}=-\frac{2a}{r^2}(r^2+a^2-\Delta_r)\ ;\  &\qquad\text{} g_{\phi\phi}=\frac{(r^2+a^2)^2-\Delta_ra^2}{r^2},
\end{alignat}
where $\Delta_r=(r^2+a^2)-2r_gr$ and $a$ is the angular momentum of the black hole. Though the Lorentz factor from Eq. (\ref{mod_lorentz}) is taken to be of the same form as for Schwarzschild spacetime, it is understood from the nature of spacetime that the velocity profiles subsume effects from black hole angular momentum.
\begin{figure}
\centering
\includegraphics[width=0.95\linewidth]{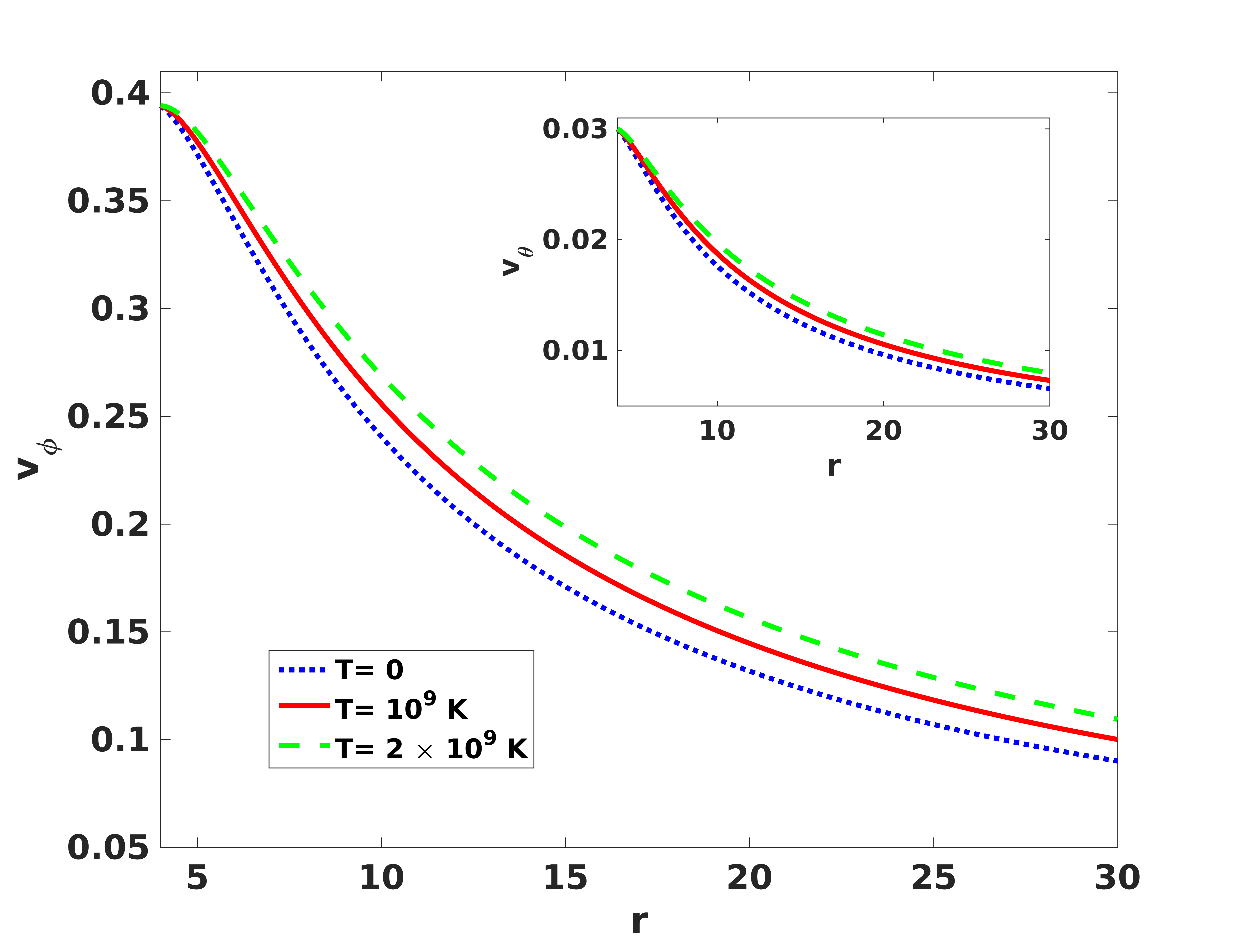}
\caption{(Color Online)Decaying solution of $v_{\phi}$ and $v_{\theta}$ in an accretion disk near a Kerr black hole for initial values of 0.394 and 0.03 respectively. Starting from r=4, three lines for both velocity profiles show decay for three different initial values of temperature from $T=0K$ (cold) to $T=2\times 10^9 K$(hot). For both velocity profiles, the decaying solution with the highest temperature decays slowest.}
\label{fig:Kerr_decay}
\end{figure}
\begin{figure}
\centering
\includegraphics[width=0.95\linewidth]{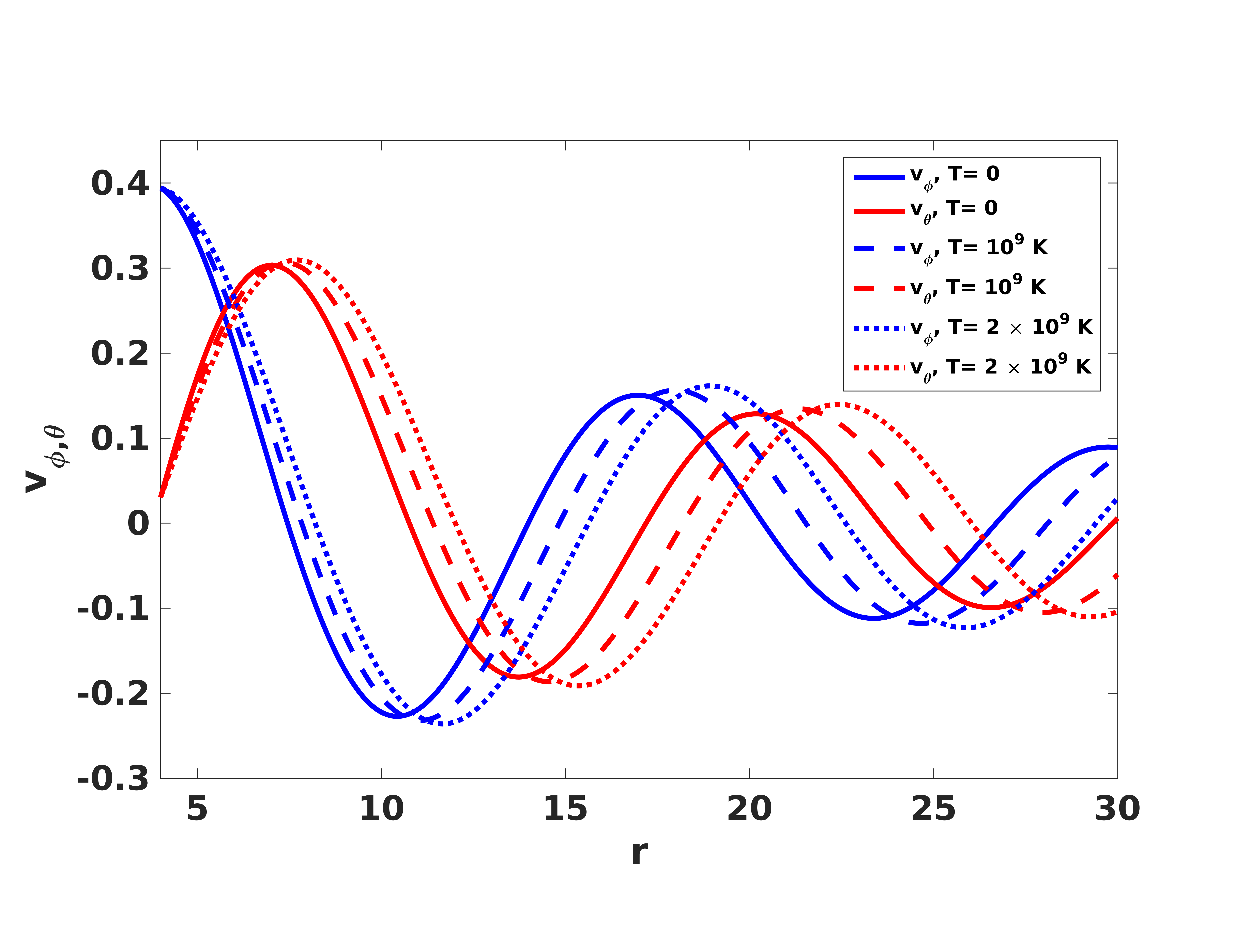}
\caption{(Color Online)Oscillatory solution of $v_{\phi}$ and $v_{\theta}$ in an accretion disk near a Kerr black hole for initial values of 0.394 and 0.03 respectively.  Starting from $r=4$, three oscillatory lines for both velocity profiles show decay for different initial temperatures $T=0K$ (solid), $T=10^9 K$ (dashed), and $T=2\times 10^9 K$(dotted). Blue lines (initially upper) and red lines (initially lower) represent $v_{\phi}$ and $v_{\theta}$ respectively. For both velocity profiles, the frequency of oscillation changes as the temperature changes.}
\label{fig:compareK_osc}
\end{figure}

Figs. (\ref{fig:Kerr_decay}) and (\ref{fig:compareK_osc}) show azimuthal and poloidal velocity profiles, respectively, in the Kerr spacetime for different temperatures starting at $r=4.0$ with $v_{\phi}(r=4)=0.394$, $v_{\theta}(r=4)=0.03$, and $a=0.99r_g$.  The reason for initializing the system at $r=4$, rather than at isco ($r=1.4545$), is discussed in Sec. \ref{disc}. The temperature profile in the accretion disk is assumed to be similar to that of a blackbody spectrum.  Both decaying and oscillatory solutions have similar features to the Schwarzschild geometry, again with the oscillatory behavior developing from a difference between $\mu_i$ and $\mu_e$. 

At this juncture, it will be interesting to understand the related physics behind the above velocity profiles representing electron-ion Beltrami states in the accretion disks around Schwarzschild and Kerr black holes. In particular, it is necessary to understand why the velocity profiles for equilibrium states in different geometries decay at different rates. One potential cause for these differences may arise from a redistribution of the total energy of the plasma in both accretion disks as it relaxes. However, a detailed account of this redistribution can only be discovered by exploring the relativistic Bernoulli condition for the plasma dynamics, which is examined next.

\section{Bernoulli Condition}\label{bc}
In GR, $\nabla_{\nu}T^{\mu \nu}=0$ does not always imply strict local energy conservation because the gravitational tidal forces work on the plasma and may increase or decrease its locally measured energy. Only when the background spacetime structure possesses some kind of symmetry, such as the static (stationary) and spherically symmetric (axisymmetry) of  Schwarzschild (Kerr) spacetimes, might the spacetime allow some form of local energy conservation. Due to the symmetry in both Schwarzschild (static and spherically symmetric) and Kerr geometry (stationary and axisymmetry), the time evolution vector field $t^\mu$, introduced in Appendix \ref{magnetofluiddynamics}, must be a Killing vector field satisfying Killing's equation $\nabla_{\mu}t_{\nu}+\nabla_{\nu}t_{\mu}=0$ \cite{Wald}. Physically, the corresponding metric along this vector field remains invariant. Also, it implies that Lie derivatives along $t^\mu$ can be considered as regular `time derivatives'. Next, since the equation of motion (\ref{finaleom}) projected along the time
evolution vector field $t^\mu$ yields the energy conservation equation, the Bernoulli condition for plasma in both geometries can be derived as follows:
\begin{eqnarray} \label{bc1}
qt^{\mu}U^{\nu}\mathcal{M}_{\mu\nu}=Tt^{\mu}\nabla_{\mu}\sigma = t^{\mu}\nabla_{\mu}H -\frac{1}{n}t^{\mu}\nabla_{\mu}p
\end{eqnarray}
with $H=\mathcal{G}/n$ denoting enthalpy per baryon. Next, with Eq. (\ref{unifiedmomentum}), (\ref{unifiedtensordef}) and $\nabla_{\mu}t_{\nu}+\nabla_{\nu}t_{\mu}=0$,  Eq. (\ref{bc1}) becomes, after some algebra,
\begin{eqnarray} \label{bc2}
&U^{\mu}\nabla_{\mu}\left(-qt^{\mu}P_{\mu}\right) + \mathcal{L}_{\vec{t}}(qU^{\mu}A_\mu) \notag \\ 
&+\frac{1}{n}\mathcal{L}_{\vec{t}}p - qA_{\mu}{L}_{\vec{t}}U^{\mu} = 0.
\end{eqnarray}
Note that we have used the generalized momentum $P^{\mu}=A^{\mu}+({mc}/{q})\mathcal{G}U^{\mu}$ defined in Eq. (\ref{unifiedmomentum}). It is obvious that the left hand side of the above equation cannot be expressed as a perfect gradient along the plasma four-velocity $U^\mu$. However, since all the Lie derivatives (time derivatives) in Eq. (\ref{bc2}) vanish for stationary and symmetric flow in the equilibrium states of plasma, the desired Bernoulli condition in covariant form manifests as
\begin{eqnarray} \label{bc3}
U^{\mu}\nabla_{\mu}\left(-t^{\mu}\tilde{P}_{\mu}\right)  = 0
\end{eqnarray}
with $\tilde{P}_{\mu}:= qP^{\mu}$.
This condition implies that the scalar quantity within the gradient, $-t^{\mu}\tilde{P}_{\mu}$, remains constant along the fluid line. However, as expected, this quantity is nothing but the time component of the generalized momentum and thereby is related to the local total energy density of plasma in its equilibrium states.

In order to express $-t^{\mu}\tilde{P}_{\mu}$ in familiar physical quantities, it is necessary to divide both sides of Eqn. (\ref{bc3}) by $-t^{\mu}\tilde{P}_{\mu}$ and rewrite the Bernoulli condition as $U^{\mu}\nabla_{\mu}\left(ln(-t^{\mu}\tilde{P}_{\mu})\right)  = 0$. Then, using $t^{\mu}=\alpha n^{\mu}+\beta^{\mu}$ from Appendix \ref{magnetofluiddynamics} to expand $-t^{\mu}\tilde{P}_{\mu}$, we find 
\begin{eqnarray} \label{BC}
ln(-t^{\mu}\tilde{P}_{\mu})  = ln\left[\alpha q A_{0} + \alpha^{2} H \Gamma \left(1-\frac{v_{\mu}\beta^{\mu}}{\alpha^2}\right) \right],
\end{eqnarray}
where $t^{\mu}A_{\mu}=A_{0} := -\alpha \Phi_E + \beta^{\mu} A_{\mu}$ (with $\Phi_E$ being the electric potential)  is the time component of the four-vector potential. While the first term of the above expansion obviously implies the electromagnetic potential energy of a plasma element in a curved background spacetime, the meaning of the second term is not clear from the way it is expressed. If the effect of the electromagnetic field was small, i.e., the first term in the above equation is much smaller than the second term (the effect of gravitational and other fields) in curved background spacetime, the meaning of the second term can become clear after re-expressing Eq. (\ref{BC}) as 
\begin{eqnarray} \label{BC1}
&ln\left[\alpha q A_{0} + \alpha^{2} H \Gamma \left(1-\frac{v_{\mu}\beta^{\mu}}{\alpha^2}\right) \right]\notag \\
&=ln\left[\alpha^{2} H \Gamma \left(1-\frac{v_{\mu}\beta^{\mu}}{\alpha^2}\right)\left(1+\frac{q A_{0} }{H \alpha \Gamma \left(1-\frac{v_{\mu}\beta^{\mu}}{\alpha^2}\right)}\right)\right] \notag \\ 
&\simeq ln (\alpha^{2}) + ln (H) + ln (\Gamma) + ln  \left(1-\frac{v_{\mu}\beta^{\mu}}{\alpha^2}\right) \notag \\ &+\frac{q A_{0} }{H \alpha \Gamma \left(1-\frac{v_{\mu}\beta^{\mu}}{\alpha^2}\right)}.
\end{eqnarray}

It is evident from the above Eq. (\ref{BC1}) that the followings take place in the Newtonian limit: $ln (\alpha^{2}) \simeq 2\Phi_{G}/c^2$ with $\Phi_{G}$ being the gravitational potential, $ln (H)$ tends towards the non-relativistic specific enthalpy, $ln (\Gamma)\simeq U^{2}/2$ for $U^{2} << 1$, $ln \left(1-\frac{v_{\mu}\beta^{\mu}}{\alpha^2}\right) \rightarrow 0$ (energy related to black hole rotation) since $\beta^{\mu} \rightarrow 0 $, and the last term goes to $-q\Phi_{E}/H$.
 
 For simple geodesic motion without electromagnetic interactions, on the other hand, a similar condition can easily be derived as  $ln (-U_{\mu}t^{\mu})= constant$ that translates into $ln (\alpha^{2}) + ln (\Gamma) + ln  \left(1-\frac{v_{\mu}\beta^{\mu}}{\alpha^2}\right)= constant.$  In the non-relativistic limit, as expected, it is the sum of the gravitational potential energy and the associated kinetic energy that remain constant.  

In principle, it is possible to determine how the total energy is redistributed by computing the relative ratios of individual to total energy from the functional form of each energy. The exact functional forms of $ln (\alpha^{2})$, $ln (\Gamma)$ and  $ln \left(1-\frac{v_{\mu}\beta^{\mu}}{\alpha^2}\right)$ can easily be determined from the metric of the background geometry. However, the functional forms of $ln (H) $ and $ A_{\mu}$ can only be determined from the structure equations of a thin disk formed from relaxed states of plasma in non-geodesic motion, which is rather involved and thus will be explored in our future endeavor. Next, we explore the related physics with the use of the above Bernoulli condition to explain the behavior of equilibrium states of plasma in black hole accretion disks.

\section{Discussion}
\label{disc}
In this paper, we have explored the general relativistic solutions of equilibrium states of a hot fluid in accretion disks within the framework of magnetofluid unification and have examined the corresponding relativistic Bernoulli condition. The non-relativistic solution has turned out to be qualitatively similar to the general relativistic solutions. However, due to the presence of spacetime curvature in the background of accreting plasma, general relativistic corrections turned out to be significant. Also, semi-relativistic temperatures of the fluid in the accretion disk change the decay rate and oscillation frequency of the velocity profile. 
\begin{figure}
\centering
\includegraphics[width=0.95\linewidth]{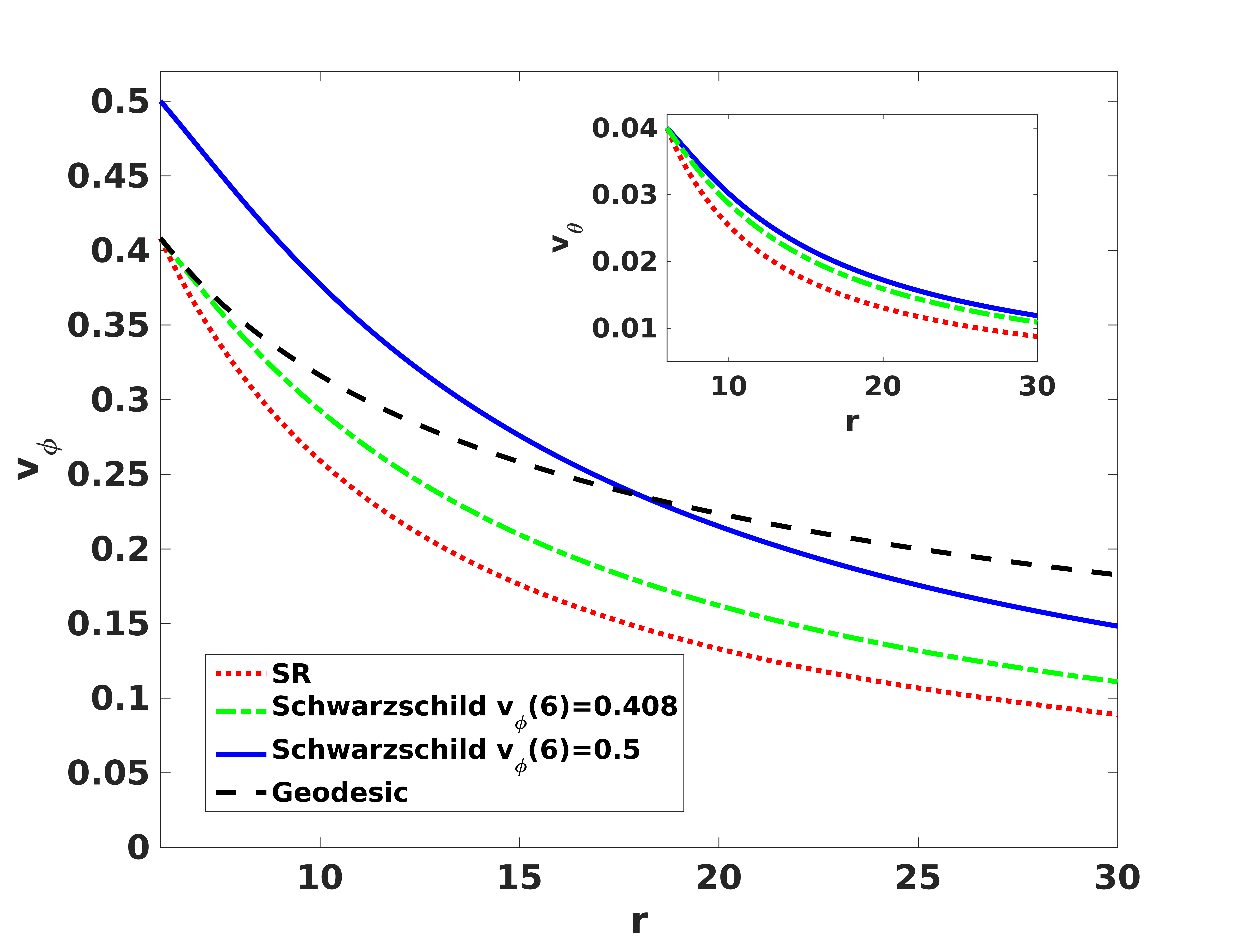}
\caption{(Color Online)Comparison between Schwarzschild and Minkowski profiles for $v_{\phi}$ and $v_{\theta}$ for cold plasma. Two azimuthal velocity profiles for nongeodesic motion with different initial values 0.5 and 0.408 (same as geodesic value at $r=6$) are shown. The initial values of the poloidal velocity profiles for Schwarzschild and Minkowski systems are kept at 0.04. }
\label{fig:compareS_Decay}
\end{figure}
\begin{figure}
\centering
\includegraphics[width=0.95\linewidth]{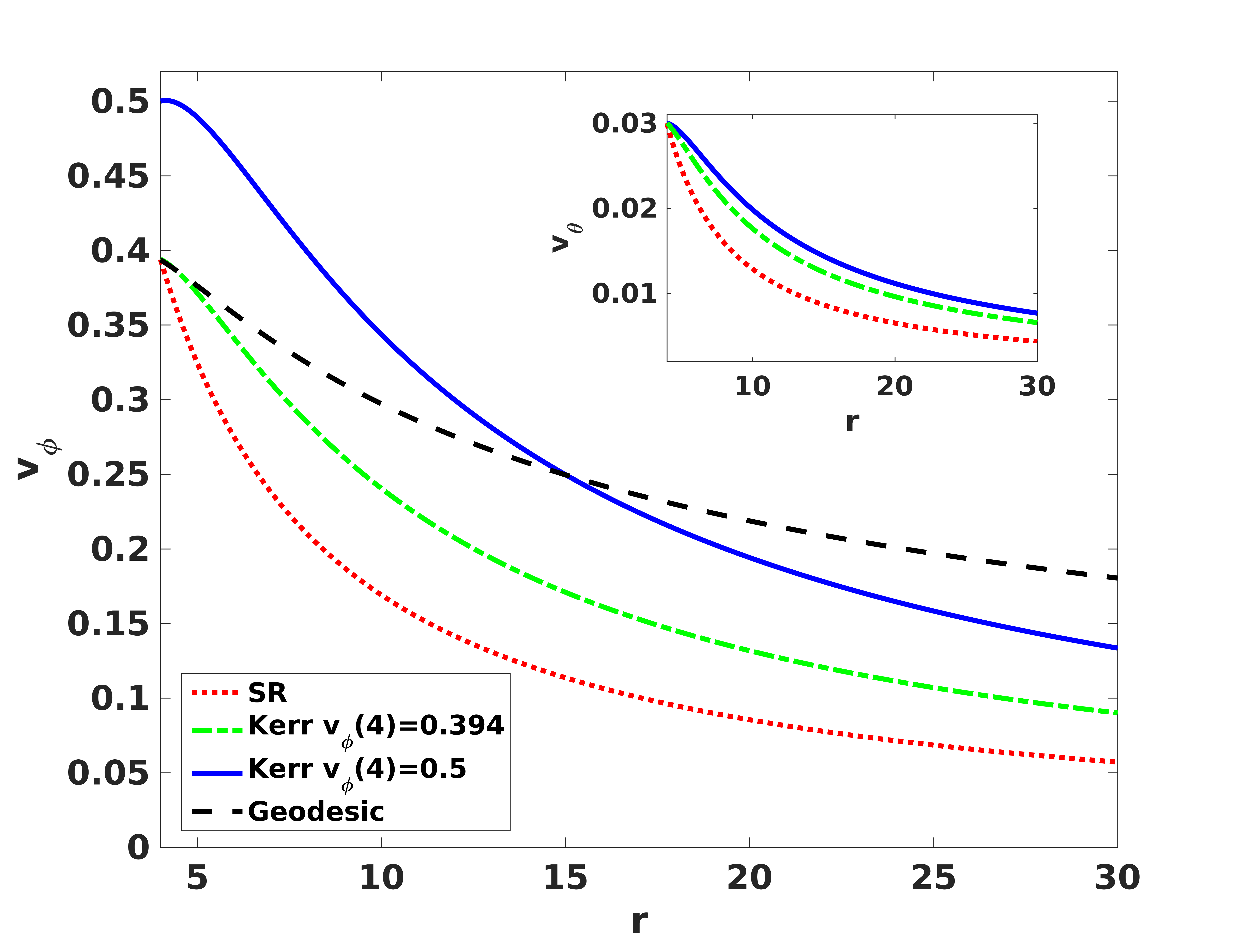}
\caption{(Color Online)Comparison between Kerr and Minkowski  profiles for $v_{\phi}$ and $v_{\theta}$ for cold plasma. Two azimuthal velocity profiles for nongeodesic motion with different initial values 0.5 and 0.394 (same as geodesic value at $r=4$) are shown. The initial values of the poloidal velocity profiles for Schwarzschild and Minkowski systems are kept at 0.03.}
\label{fig:compareK_Decay}
\end{figure}

Figs.(\ref{fig:compareS_Decay}) and (\ref{fig:compareK_Decay}) show comparative decaying ($\epsilon=0$) velocity profiles for accretion disk plasmas in Schwarzschild and Kerr geometries respectively. Each set contains solutions for two different initial conditions, and also for Minkowski spacetime (under the appropriate special relativistic limits). Both of these figures present, additionally,  the azimuthal velocity profile $v_{\phi}$ for geodesic motion in a gravitational field. The general expression for $v_{\phi}$ in geodesic motion is \cite{bhattacharjee2015novel}
\begin{equation}\label{vgeodesic}
v_{\phi}=r \ \frac{d\phi}{dt}=r \ \frac{-g_{t\phi,r}+\sqrt{(g_{t\phi,r})^2-g_{tt,r}g_{\phi\phi,r}}}{g_{\phi\phi,r}},
\end{equation}
where we assumed that fluid elements are moving in Keplerian orbits. Solving for geodesic motion in the spacetime of a Schwarzschild black hole yields a very simple solution $v_{\phi}=1/\sqrt{r}$. We need to resort to numerical means to find the geodesic velocity profile for a Kerr black hole. The calculated velocity profile, along with the associated Lorentz factor, for a Kerr black hole increases at very small radii (progressing to larger $r$) before settling into its decay \cite{bhattacharjee2015novel}. To isolate the large-scale decaying behavior of the equilibrium states, we initialized our Kerr geometry profiles at $r=4$, outside of this anomalous region. In order to avoid redundancy, we did not include a comparative plot on oscillating solutions in this section. It must be emphasized that a nonzero $v_{\theta}$ is required for the non-geodesic solutions (also presented in Figs. (\ref{fig:compareS_Decay}) and (\ref{fig:compareK_Decay})), each exhibiting a similar decay rate to its $v_{\phi}$ counterpart. Of particular interest are the velocity profiles for Minkowski spacetime, which exhibit faster decay rates than the general relativistic systems. 

At this point, it would be desirable to substantiate our motivations for resorting to the above geodesic velocity profiles for comparison. First, as discussed in section \ref{GR}, turning off thermodynamics and electromagnetism in non-geodesic motion of plasma presents us with two distinct systems but with similar mathematical structure: i) non-interacting neutral fluid (dust particles) in geodesic motion in the accretion disk, and ii) discrete plasma particles in geodesic motion. An analysis of the former system (the fluid model), however, encounters  the same issues with arbitrariness in initial values as discussed in section \ref{SC} and, therefore, its solution cannot be used as a standard velocity profile for comparison with our equilibrium hot charged fluid. On the contrary, the analysis of the latter system, though void of properties of interacting continuous media, offers us exact velocity profiles determined exclusively from metric elements and their radial derivatives as shown in Eq. (\ref{vgeodesic}). Therefore, the comparison of these known velocity profiles to the non-geodesic ones, along with the corresponding relativistic Bernoulli conditions, unravels the associated physics concealed in an N-particle system (a hot charged fluid in the accretion disk of a black hole) interacting via thermodynamics and electromagnetism. 

Now, the comparative velocity profiles in Figs. (\ref{fig:compareS_Decay}) and (\ref{fig:compareK_Decay}) can be understood by examining Eq. (\ref{BC1}) derived in the discussion on the Bernoulli condition. First, it is obvious that the  inclusion of general relativistic corrections has slowed down the decay rate (or growth rate) of velocity profiles significantly for both background geometries. The plasmas further away from the black holes are moving faster compared to its motion in flat (Minkowski) spacetime. Second, for both geometries, the non-geodesic velocity profiles would decay faster than the corresponding geodesic velocity profiles, even though the same initial velocity is assumed for both cases. We included an additional initial value solution to better compare the decay rates with geodesic motion. Though the two corresponding dynamic systems are intrinsically different, such behavior of faster decay rates arises from the possibility that the kinetic energy of plasma in non-geodesic motion is redistributed into electromagnetic, kinetic, gravitational and thermal energies of the system. This results in a lower kinetic energy for non geodesic motion, but the geodesic kinetic energy is transformed into only gravitational energy as the radius of the circular orbit increases.

Also, it is obvious from  Eq. (\ref{BC1}) that the term $ln  \left(1-\frac{v_{\mu}\beta^{\mu}}{\alpha^2}\right)$ signifies a  higher total energy for  plasmas in Kerr as compared to Schwarzschild geometry. Hence,  though the realm of stable circular orbits are different, the velocity profiles in Kerr geometry will decay slower than that in Schwarzschild geometry, which can be seen from the corresponding radial derivatives of the velocity profiles. The plasmas surrounding a Kerr blackhole can maintain stable circular orbits at distances closer than that of the closest stable orbits in Schwarzschild geometry, which can be attributed to the rotation in Kerr spacetime \cite{2013A&A551A4P,bhattacharjee2015novel}. The Bernoulli condition, Eqn. (\ref{BC1}), clearly shows that, for non-geodesic plasma flow in the accretion disks for both geometries, the plasma is endowed with additional energies from thermodynamics and electromagnetism, thus displaying a rich interplay of plasma dynamics and GR.

\begin{figure}[t]
\centering
\includegraphics[width=1.02\linewidth]{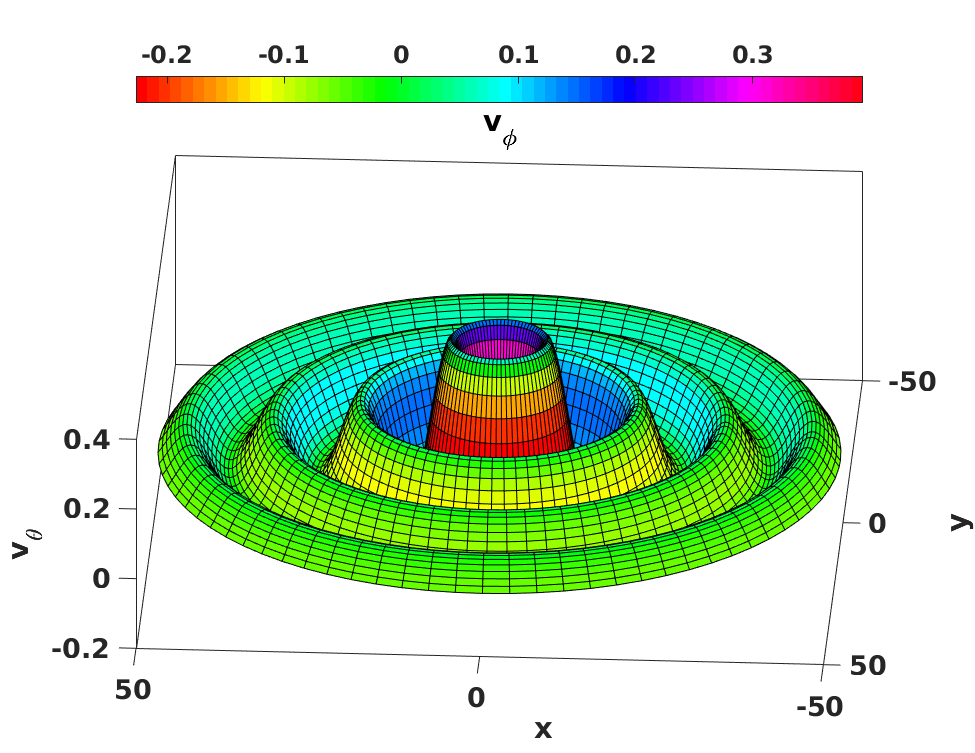}
\caption{(Color Online)Poloidal velocity for the Kerr oscillating ($\epsilon=0.5$) solution in the plane of the disk. $v_{\phi}$ is given by the color scheme whereas $v_{\theta}$ represents height in the plot. The axes from $-40$ to $40$ depict the length scale in the plane of the disk in units of $r_g$.}
\label{fig:3d}
\end{figure}

On the other hand, the oscillating solutions displayed in Figs. (\ref{fig:classical}), (\ref{fig:compareS_osc}), and (\ref{fig:compareK_osc}) show that $v_{\theta}$ oscillates with the same amplitude as $v_{\phi}$. The presence of such large azimuthal velocities calls into question the disk structure we have assumed to describe our system. To better visualize this system (in comparison to the decaying solution), we show a 3D rendering of the Kerr solution ($T=0$ and $\epsilon=0.5$) in Fig. (\ref{fig:3d}). The largest velocities out of the plane are very close to the black hole. These could account for steady plasma winds from the disk that could fuel ambipolar relativistic jets. This motion outside of the plane is beyond the scope of this current paper; a more detailed study of disk-jet physics would be required to understand more precisely the mechanisms at play. 

We recall that this oscillatory behavior is traced back to the difference $\epsilon=\mu_i-\mu_e$ between the Lagrange multipliers, which in turn depends upon each species' conserved generalized helicity. The associated helicities have contributions from both the electromagnetic and flow fields. As the mass of the electron is negligible, there will be a difference between the two helicities:
\begin{eqnarray}
\mathcal{H}_e=<\vec{B}\cdot\vec{A}>, \label{hel_elec}\\
\mathcal{H}_i=<(\vec{B}+\frac{m_ic}{q}\vec{\nabla}\times(\vec{\mathcal{G}\Gamma_iv_i}))\cdot(\vec{A}+\frac{m_ic}{q}\mathcal{G}\vec{v_i})>. \label{hel_ion}
\end{eqnarray}
 The microscopic phenomenon related to the corresponding difference in Lagrange multipliers manifests itself in the large scale oscillatory velocity profile. With changing $\epsilon$, the frequency of the macroscopic oscillating velocity profile changes.

\section{Conclusion}

The relaxation of a two species magnetofluid in an accretion disk is described in the presence of gravity. Relaxation is achieved by a constrained energy minimization; the constraints are provided by the demands of preserving the system invariants as the dynamics evolves. For the Hall MHD model of the electron-ion plasma, the invariants are the magnetic and the generalized helicities.  

Such relaxed states seem to be accessible to accretion disk plasmas that are subject to various forces including the ones stemming from thermodynamic potential gradients, curvature effects, shear flows etc, as evident from the corresponding Bernoulli condition. In this paper, we have considered a simple 1D problem in the $\theta=\pi/2$ plane of an accretion disk to delineate the GR effects with approximate analytical  and numerical methods. 

We find that these equilibria are  significantly different from both their NR counterparts; the plasma dynamics is also very different from purely geodesic motion. Effects arising from curved spacetime, semi-relativistic temperature and blackhole rotation give a rich and more restrictive structure to the equilibrium configuration. The added physics in non-geodesic motion forces the velocity and field profiles into steeper decay rates than that in geodesic motion, supported by the analysis of the Bernoulli condition. 

The oscillatory solutions reveal that the accretion disk will have regions of significant azimuthal flow. These states, in principle,  could be responsible for a collimated jet structure emanating from a black hole,  as the plasma elements can follow the vortical field lines emanating out of the disk \cite{shatashvili2011generalized}. 

This paper, however, constitutes a conceptual and preliminary but, hopefully, an insightful investigation of  plasmas in accretion disks surrounding very compact objects. The real accretion disks have finite width  with density and temperature varying along the vertical direction \cite{skadowski2011relativistic}. A full treatment will require at least a 2D solution in different complex geometries \cite{25270,7726,25339,25340,14718}.

\begin{appendix}
	
	\section{3+1 Dynamics of GravitoMagnetofluid }\label{magnetofluiddynamics}

	The approach chosen for the 3+1 splitting selects a family of foliated fiducial 3-dimensional hypersurfaces (slices of simultaneity) $\Sigma_{t}$ labelled by a parameter $t = constant$ in terms of a time function on the manifold. Furthermore, we let $t^{\mu}$ be a timelike vector whose integral curves intersect each leaf $\Sigma_{t}$ of the foliation precisely once and which is normalized such that $t^{\mu}\nabla_{\mu}t = 1$. This $t^{\mu}$ is the `evolution vector field' along the orbits of which different points on all  $\Sigma_{t} \equiv \Sigma$ can be identified. This allows us to write all space-time fields in terms of $t$-dependent components defined on the spatial manifold $\Sigma_{t}$. Lie derivatives of space-time field along $t^{\mu}$ are identified with ``time derivatives" of the spatial fields since Lie derivatives reduce to partial time derivative for an adapted coordinate system $t^{\mu}=(1,0,0,0)$.
	
	Moreover, since we are using the Lorentzian signature, the vector field $t^{\mu}$ is required to be future directed. Let us decompose $t^{\mu}$ into normal and tangential parts with respect to $\Sigma_{t}$ by defining the lapse function $\alpha$ and the shift vector $\beta^{\mu}$ as $t^{\mu}=\alpha n^{\mu}+\beta^{\mu}$ with $\beta^{\mu}n_{\mu} = 0$, where  $n^{\mu}$ is the future directed unit normal vector field to the hypersurfaces $\Sigma_{t}$. More precisely, the natural timelike covector $n_{\mu}=(-\alpha,0,0,0)=-\alpha\nabla_{\mu}t$  is defined to obtain $n^{\mu}=({1}/{\alpha},-\beta^{\mu}/\alpha)$ which satisfy the normalization condition $n^{\mu}n_{\mu}=-1$. Then, the space-time metric $g_{\mu\nu}$ induces a spatial metric $\gamma_{\mu\nu}$ by the formula $\gamma_{\mu\nu}=g_{\mu\nu}+n_{\mu}n_{\nu}$. Finally, the 3+1 decomposition is usually carried out with the projection operator $\gamma^{\mu}\ _{\nu}=\delta^{\mu} \ _ {\nu}+n^{\mu}n_{\nu}$, which satisfies the cond!
 ition $n^{\mu}\gamma_{\mu\nu}=0$. Also, the acceleration is defined as $a_{\mu}=n^{\nu}\nabla_{\nu}n_{\mu}$. \\
	
	Now, with the above foliation of space-time, the space-time metric  takes the following canonical form \cite{MTW}
	\begin{equation}\label{canonicalmetric}
	ds^2=-\alpha^2dt^2+\gamma_{ij}(dx^i+\beta^i dt)(dx^j+\beta^j dt),
	\end{equation}
	and it immediately follows that, with respect to an Eulerian observer, the Lorentz factor turns out to be
	\begin{equation}\label{lorentzfactor}
	\Gamma=\left[\alpha^2-\gamma_{ij}(\beta^{i}\beta^{j}+2\beta^{i}v^{j}+v^{i}v^{j})\right]^{-1/2},
	\end{equation}
	satisfying $d\tau = dt/\Gamma$, where $v^i$ is the $i$th component of fluid velocity $\vec{v}=d\vec{x}/dt$.  Then the decomposition for the 4-velocity is \cite{asenjo2013generating}
	\begin{equation}\label{fourvelocity}
	U^{\mu}=\alpha \Gamma n^{\mu}+\Gamma\gamma^{\mu} \ _{\nu}v^{\nu},
	\end{equation}
	with $n_{\mu}U^{\mu}=-\alpha \Gamma$.

	Now, since our unified anti-symmetric field tensor $\mathcal{M}^{\mu\nu}$ is constructed from the antisymmetric tensors $F^{\mu\nu}$ and $D^{\mu\nu}$, we apply the ADM formalism of electrodynamics presented in \cite{thorne1982electrodynamics,thorne1986black,MTW,Wald} to define the generalized electric and magnetic field, respectively, as 
	\begin{alignat}{2}\label{generalemdef}
	\xi^{\mu}=n_{\nu}\mathcal{M}^{\mu\nu}\ ;\  &\qquad\text{}\qquad \Omega^{\mu}=\frac{1}{2}n_{\rho}\epsilon^{\rho\mu\sigma\tau}\mathcal{M}_{\sigma\tau},
	\end{alignat}
	and thus express the unified field tensor
	\begin{equation}\label{generalfieldtensordef}
	\mathcal{M}^{\mu\nu}=n^{\mu}\xi^{\nu} - n^{\nu}\xi^{\mu}-\epsilon^{\mu\nu\rho\sigma}\Omega_{\rho}n_{\sigma}.
	\end{equation}
	We remind the reader that the generalized magnetic field and the generalized vorticity are essentially synonymous. Using the definition of the unified field tensor $\mathcal{M}^{\mu\nu}$,  the expressions of 3D generalized electric and magnetic fields turn out to be
	
	\begin{align}\label{generalefield}
	\vec{\xi}= \ & \vec{E}-\frac{m}{q}(1+\lambda f_{m}(R)-\lambda RF_{m}(R))\vec{\nabla}(\alpha \mathcal{G}\Gamma)\notag \\
	&-\frac{m}{q}\lambda F_{m}(R)\vec{\nabla}(\alpha \mathcal{G}R\Gamma)\notag \\
	& -\frac{m}{q}(1+\lambda f_{m}(R))\left[2\underline{\underline{{\sigma}}}\cdot(\mathcal{G}\Gamma\vec{v})+\frac{2}{3} \Theta\mathcal{G}\Gamma\vec{v}\right]\notag \\
	& -\frac{m}{q\alpha}(1+\lambda f_{m}(R)-\lambda RF_{m}(R))\left(\mathcal{L}_t(\mathcal{G}\Gamma\vec{v}) -\mathcal{L}_{\vec{\beta}}(\mathcal{G}\Gamma\vec{v})\right)\notag  \\
	& -\frac{m}{q\alpha}\lambda F_{m}(R)\left(\mathcal{L}_t(\mathcal{G}R \Gamma\vec{v}) -\mathcal{L}_{\vec{\beta}}(\mathcal{G}R \Gamma\vec{v})\right);
	\end{align}
	\begin{align}\label{generalbfield}
	\vec{\Omega}=\vec{B}+\frac{m}{q}(1+\lambda f_{m}(R)-\lambda RF_{m}(R)) \vec{\nabla}\times(\mathcal{G}\Gamma\vec{v})\notag \\
	+\lambda F_{m}(R) \frac{m}{q}\vec{\nabla}\times(R\mathcal{G}\Gamma\vec{v}),
	\end{align}
	where $\underline{\underline{{\sigma}}} = \sigma_{\mu}^{\nu}$ and $\Theta$ are, respectively,   the shear and expansion of the congruence, defined as $\sigma_{\alpha\beta} = \gamma_{\alpha}^{\mu}\gamma_{\beta}^{\nu}\nabla_{(\mu} n_{\nu)}- \frac{1}{3}\theta \gamma_{\mu\nu}$ and $\Theta =\nabla_{\mu}n^{\mu} $. We have also used the relation $\nabla_{\mu} n_{\nu} = -a_{\nu}n_{\mu} + \sigma_{\alpha\beta} + \frac{1}{3}\theta \gamma_{\mu\nu}$ to derive (\ref{generalefield}). Finally, we used $\mathcal{L}$ to denote Lie derivative.

\end{appendix}

\bibliographystyle{unsrt}
\bibliography{ref}
\end{document}